\documentclass{moriond}

\def\Journal#1#2#3#4{{#1} {\bf #2}, #3 (#4)}

\newcommand{\aap}{Astron. Astrophys.}   %
\newcommand{\jcap}{J. Cosmol. Astropart. Phys.}   %
\newcommand{\mnras}{Mon. Not. R. Astron. Soc.}   %

\def\be{\begin{equation}}
\def\ee{\end{equation}}
\def\bea{\begin{eqnarray}}
\def\eea{\end{eqnarray}}

\newcommand{\Photo}{\includegraphics[height=35mm]{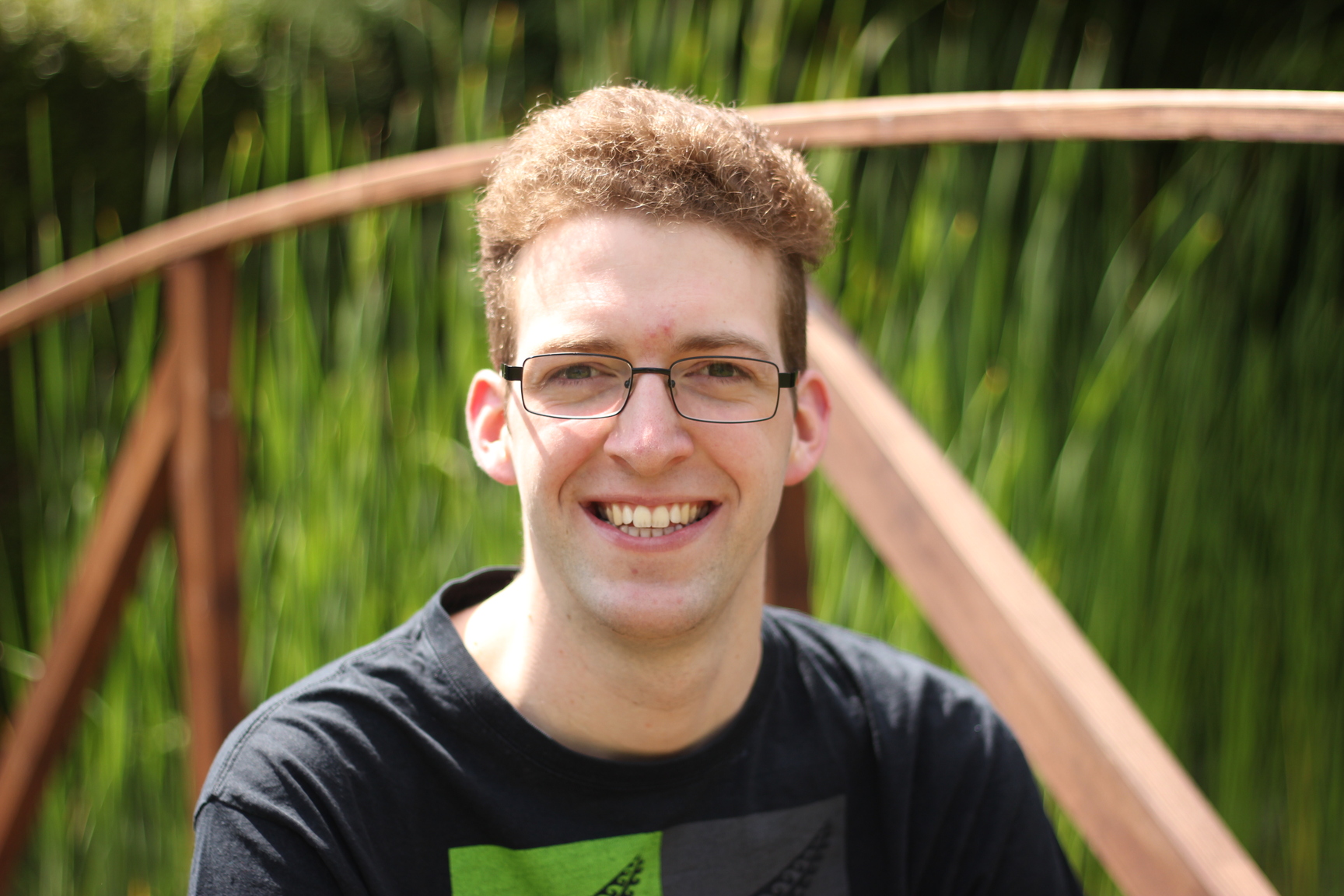}}
\usepackage{amssymb}
\usepackage{xcolor}

\begin{document}
\vspace*{4cm}
\title{\textcolor{gray}{Contribution to the 2022 Cosmology session of the 56th Rencontres de Moriond}\vspace{1em}\\What it takes to measure Reionization with Fast Radio Bursts}

\author{Stefan Heimersheim}

\address{Institute of Astronomy, University of Cambridge,\\
Madingley Road, Cambridge CB3 0HA, UK}

\vspace{-2em}%
\maketitle\abstracts{
Fast Radio Bursts (FRBs) are recently discovered extra-galactic radio transients which are now used as novel cosmological probes. We show how the Bursts' Dispersion Measure can model-independently probe the history of Hydrogen reionization. Using a FlexKnot free-form parameterization to reconstruct the reionization history we predict an 11\% accuracy constraint on the CMB optical depth, and 4\% accuracy on the midpoint of reionization, to be achieved with 100 FRBs originating from redshifts $z>5$.}

\section{Introduction}
Fast Radio Bursts (FRBs) are short ($\sim 1\,$ms) bright bursts originating from outside of our galaxy.
What makes FRBs an excellent cosmological probe is that the signal is dispersed by the free electrons in the Universe, most significantly the intergalactic medium (IGM) but also the Milky Way and host galaxy.
Measuring this dispersion (DM) allows us to probe the integrated electron column density
and thus probe distance or ionization state of the IGM. 

The idea to constrain reionization, and in particular the CMB optical depth $\tau$, with FRBs from high redshift ($z>5$) has been proposed as early as 2016 \cite{F16} and has gained traction as thousands of FRBs have been discovered since.
Even though no FRBs have been observed from the Epoch of Reionization, forecasts \cite{Hashimoto} expect $\sim 100\mathrm{\ FRBs/day/sky}$ from $z>6$ to be observed with SKA.

\section{Method}
In our work \cite{Heimersheim} we address a problem of reionization constraints derived from FRBs, namely that the
results are intrinsically sensitive to the shape of the reionization history $x_i(z)$.
Because the dispersion measure gives an integral constraint on $x_i(z)$, conversion into the
optical depth or ionized fraction depends on the model or parameterization assumed for  $x_i(z)$.
Besides pathological examples (e.g. sharp dips and peaks in the ionization history) this effect is relevant in practice -- for example the commonly used \textit{tanh} step function is an inadequate fit to the real ionization history and can deliver inaccurate results for realistic reionization scenarios.

\begin{figure}[h]
    \centering
    \includegraphics[height=1.5in]{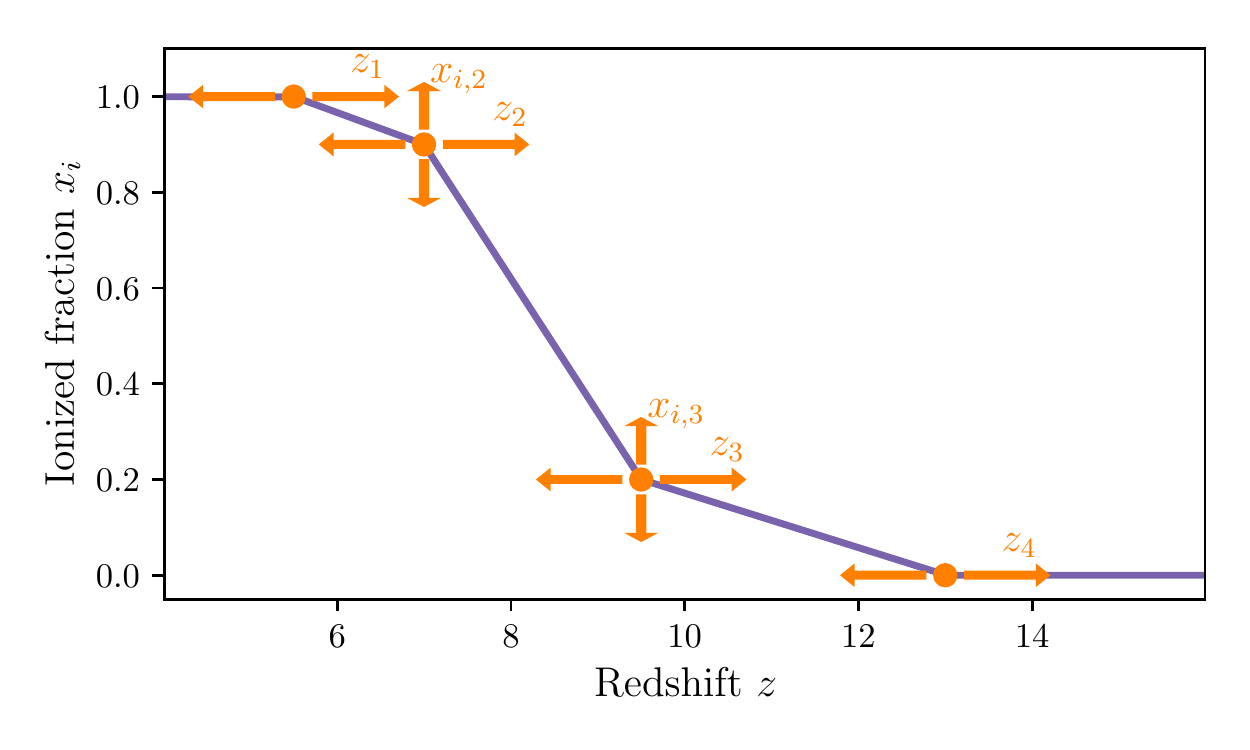}
    \hspace{0.1\textwidth}
    \includegraphics[height=1.5in]{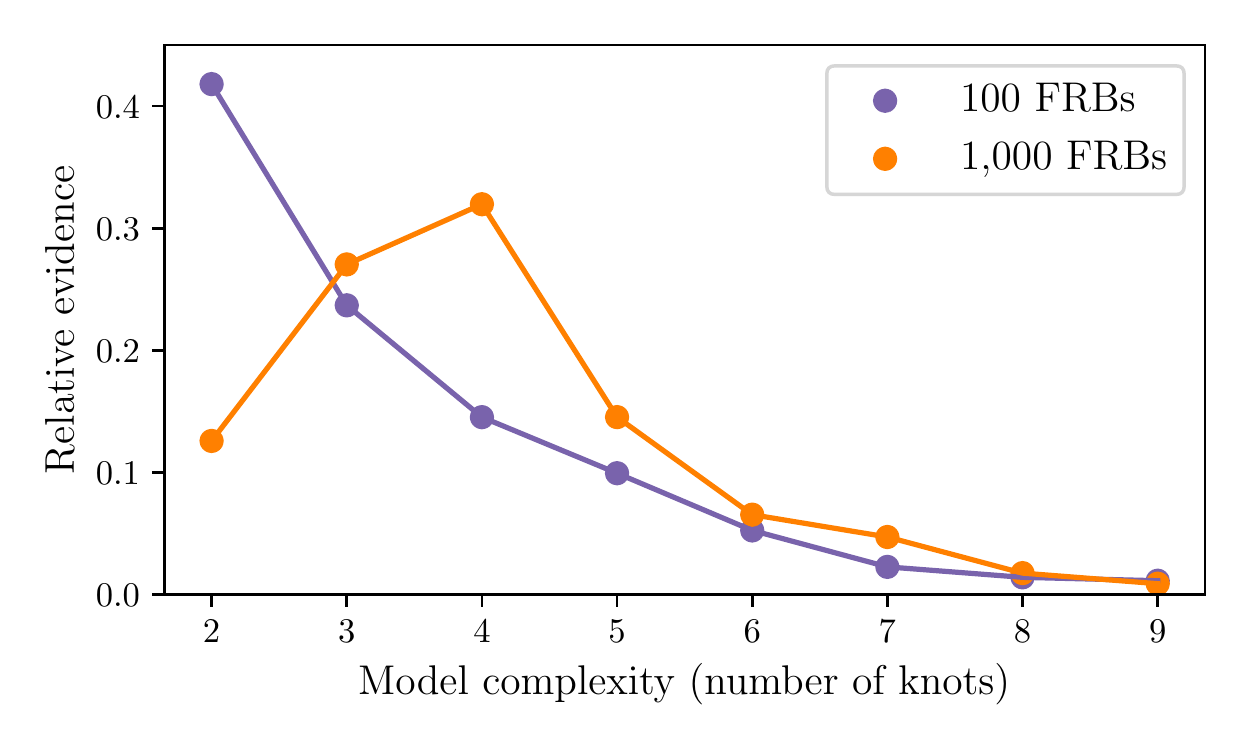}
    \caption{\textbf{Left:} Illustration of the FlexKnot parameterization. \textbf{Right:} Evidence as function of knot number.}
    \label{fig1}
\end{figure}

Ideally we want to marginalize over all possible reionization histories i.e. derive constraints while averaging over all shapes, weighted by their likelihood.
This is exactly what we propose to do with the ``model-independent'' FlexKnot method \cite{Millea} -- we parameterize  $x_i(z)$ as an interpolation function between a (varying) number of interpolation knots, as shown in Figure \ref{fig1} (left). This allows the algorithm to capture reionization histories of arbitrary complexity. We then marginalize over the number of knots by averaging all cases weighted by their Bayesian evidence as shown in Figure \ref{fig1} (right).
The evidence tends to fall when the number of knots exceeds the complexity warranted by the data, naturally creating an Occam's razor effect.

\section{Results}

\begin{figure}[h]
    \centering
    \includegraphics[height=1.5in]{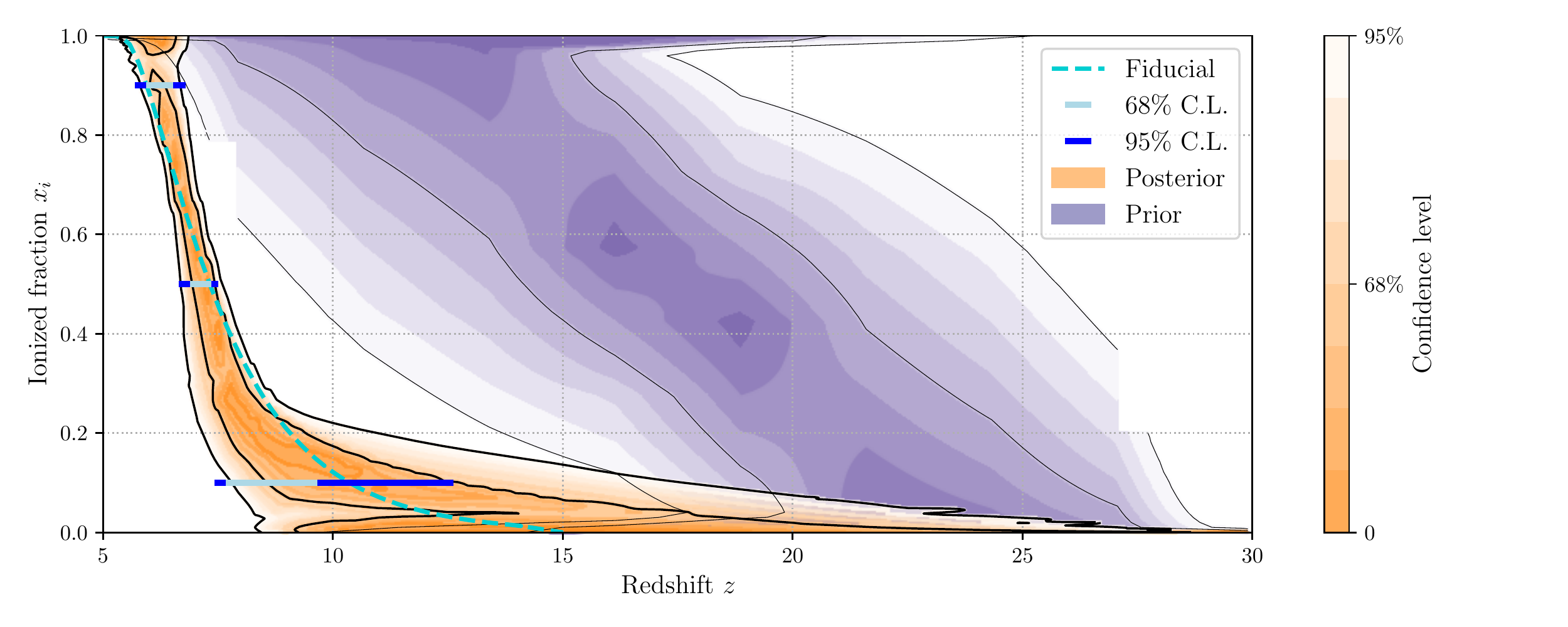}
    \includegraphics[height=1.5in]{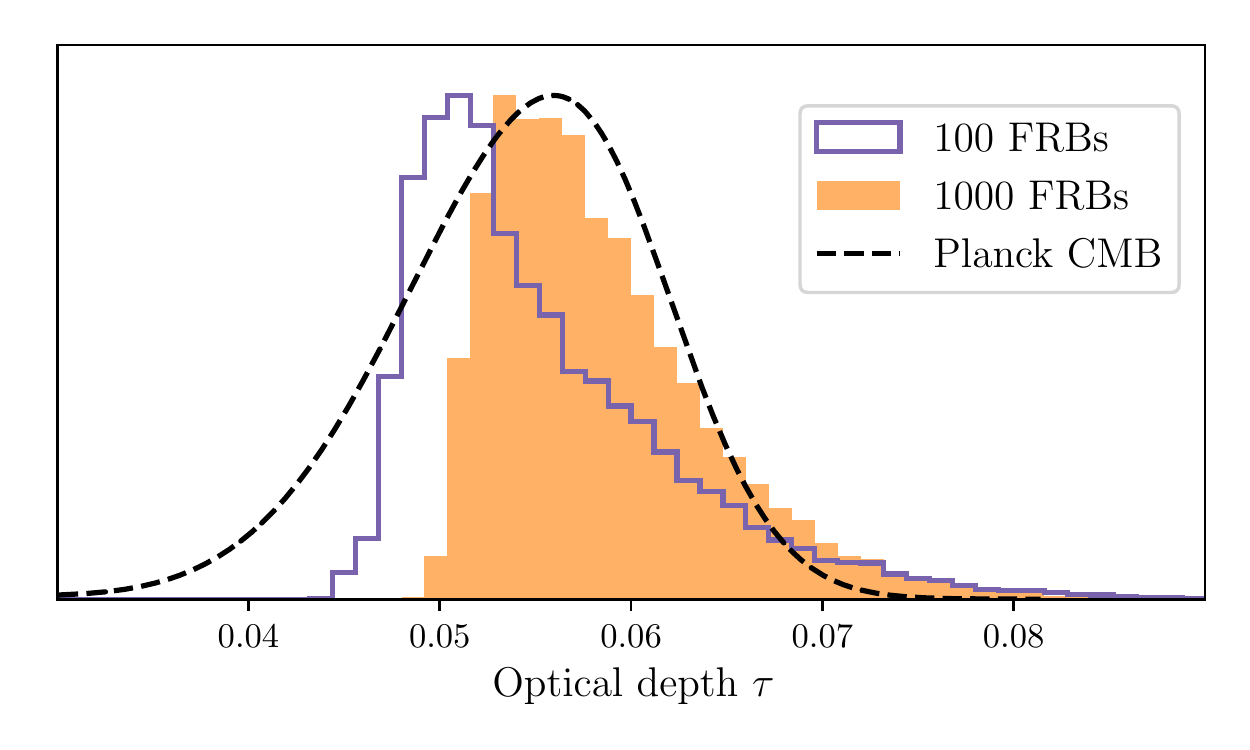}
    \caption{\textbf{Left:} Constraints on reionization history for 1000 FRBs. \textbf{Right:} Posterior of optical depth $\tau$.
    }
    \label{fig2}
\end{figure}

Having sampled all parameters and reionization histories we can derive the Bayesian posterior distribution for $x_i(z)$ as shown in Figure \ref{fig2} (left). The countours show the 68\% and 95\% confidence interval of $x_i$ for every redshift $z$. The posterior provides tight constraints on $x_i(z)$ and agree well with the input fiducial history
which was used to generate the data.

In addition we can constrain any derived quantity such as the midpoint of reionization (central blue lines) or the optical depth $\tau$ (Figure \ref{fig2}, right) by computing the quantity for every sampled history and then marginalizing over parameters. The expected midpoint constraints are competitive with current quasar constraints, and the optical
depth measurement can improve current limits from the CMB.
More details on both the method and results can be found in our paper \cite{Heimersheim}.

\end{document}